\documentclass[aps,prl,twocolumn,superscriptaddress,amsfont,graphicx,nofootinbib,preprintnumbers]{revtex4}%
\usepackage{color,graphicx,epsfig}
\usepackage{ifpdf}
\usepackage{amsmath}
\usepackage{bm}
\usepackage{color}
\usepackage[english]{babel}
\usepackage{graphicx}%
\usepackage{amsfonts}%
\usepackage{amssymb}
\usepackage{braket}
\usepackage{hyperref}
\usepackage{enumerate}

\definecolor{nicered}{rgb}{0.7,0.1,0.1}
\definecolor{nicegreen}{rgb}{0.1,0.5,0.1}
\hypersetup{colorlinks,citecolor= nicegreen,linkcolor= nicered}

\newcommand{\beq}{\begin{equation}}
\newcommand{\eeq}{\end{equation}}
\newcommand{\bea}{\begin{eqnarray}}
\newcommand{\eea}{\end{eqnarray}}

\definecolor{Red}{rgb}{1.,0.,0.}

\newcommand{\BR}{{\rm BR}}

\def\mysection#1{{{\bf #1}.~}}
\def\OMIT#1{}

\begin{document}

\def\Cincy{Department of Physics, University of Cincinnati, Cincinnati, Ohio 45221,USA}
\def\CERN{CERN Theory Division, CH-1211, Geneva 23, Switzerland}
\def\Weiz{Department of Particle Physics and Astrophysics, Weizmann Institute of Science, Rehovot 7610001, Israel}
\def\Argonne{High Energy Physics Division, Argonne National Laboratory, Argonne, IL 60439, USA}
\def\Northwestern{Department of Physics \& Astronomy, Northwestern University, Evanston, IL 60208, USA}


\title{An Exclusive Window onto Higgs Yukawa Couplings}

\author{Alexander L. Kagan}
\email[Electronic address:]{kaganal@ucmail.uc.edu}
\affiliation{\Cincy}

\author{Gilad Perez}            
\email[Electronic address:]{gilad.perez@cern.ch}
\affiliation{\CERN}
\affiliation{\Weiz}

\author{Frank Petriello}     
\email[Electronic address:]{f-petriello@northwestern.edu}
\affiliation{\Argonne}
\affiliation{\Northwestern}

\author{Yotam Soreq}            
\email[Electronic address:]{yotam.soreq@weizmann.ac.il}
\affiliation{\Weiz}

\author{Stoyan Stoynev}            
\email[Electronic address:]{stoyan.stoynev@cern.ch}
\affiliation{\Northwestern}

\author{Jure Zupan} 
\email[Electronic address:]{zupanje@ucmail.uc.edu} 
\affiliation{\Cincy}

\date{\today}
\begin{abstract}
We show that both flavor-conserving and flavor-violating Yukawa couplings of the Higgs boson to first- and second-generation quarks can be probed by measuring rare decays of the form $h \to MV$, where $M$ denotes a vector meson and $V$ indicates either $\gamma,W$ or $Z$.  We calculate the branching ratios for these processes in both the Standard Model and its possible extensions. We discuss the experimental prospects for their observation.  The possibility of accessing these Higgs couplings appears to be unique to the high-luminosity LHC and future hadron colliders, providing further motivation for those machines.
\end{abstract}

\maketitle

\mysection{Introduction} \label{sec:intro}
The discovery of a Higgs-like boson by the ATLAS and CMS collaborations~\cite{:2012gk,:2012gu} ushered in a new era of exploration in high-energy physics driven by the desire to understand the properties of this new state.  Current measurements only give information on the Higgs couplings to gauge bosons and third-generation fermions.  In the well-measured decay modes, $h\to \gamma\gamma$, $WW$ and $ZZ$,  the measured couplings agree with the Standard Model (SM) values at the $20-30$\% level~\cite{atlascoup,cmscoup}. Initial evidence in the $h\to \tau\tau$ channel also indicates no deviation from the SM value~\cite{Chatrchyan:2014nva}.

In contrast, the couplings of the Higgs to the first- and second-generation fermions are only weakly constrained by the inclusive Higgs production cross sections.  At the same time they can receive large modifications in beyond-the-SM theories, making them interesting experimental targets.  The ATLAS and CMS collaborations have studied the possibility of measuring the Higgs couplings to muons at the high-luminosity LHC (HL-LHC) with encouraging results  \cite{ATLAS-CONF-2013-010,CMS-PAS-HIG-13-007,Dawson:2013bba}. However, the question of whether the Higgs couplings to the other light fermions can be directly accessed is left completely open. 

In this Letter we lay out a program to  measure enhanced Higgs couplings to light quarks using the exclusive decays $h \to MV$, where $M$ denotes a vector meson and $V$ denotes either a $\gamma, W$ or $Z$.  The possibility of using $h \to J/\psi\gamma$ to probe the flavor-diagonal Higgs coupling to charm quarks was recently pointed out in~\cite{Bodwin:2013gca}.  This coupling can also be accessed using charm-tagging techniques~\cite{Delaunay:2013pja}.  The modes studied here, on the other hand, allow access to Higgs couplings that are impossible to directly determine in other ways.  For example, the $h \to \phi\gamma$ mode considered here allows direct access to the flavor-diagonal coupling of the Higgs to the strange quark.  The $h\to \rho\gamma, \omega\gamma$ modes probe the Higgs couplings to up and down quarks, while the $h \to K^{*0}\gamma,D^{*0}\gamma,B^{*0}\gamma,B_s^{*0}\gamma$ modes probe the off-diagonal Yukawa couplings of the Higgs.  We discuss the signatures of the above rare radiative processes, and also comment on $h\to M W,\, M Z$ decays, e.g. $h\to  B^{(*)+} W^-$.  Probes of the electroweak couplings of the Higgs via $h\to MZ, MW$ decays have been discussed in \cite{Isidori:2013cla}.

These rare decays are only accessible at the HL-LHC and future high-energy colliders, due to their small branching ratios. We note that the predicted event rates at planned $e^+ e^- $ facilities are too small.  This strengthens the motivation for future hadron colliders.

\mysection{Theoretical framework}
We first consider the constraints on the Higgs Yukawa couplings coming from the inclusive Higgs production rate at the LHC.  In our analysis we assume that there is only one Higgs scalar with mass $m_h\simeq125.7\,$GeV, which is a singlet of the custodial symmetry preserved by electroweak symmetry breaking and has CP conserving couplings. The effective Lagrangian used in our analysis is
\begin{align} 
	\label{eq:Leff}
	{\cal L}_{\rm eff} 
&=	-\sum_{q=u,d,s} \bar \kappa_q \frac{m_b}{v} h \bar{q}_L q_R - \sum_{q\ne q'} \bar \kappa_{qq'} \frac{m_b}{v} h \bar{q}_L q'_R + h.c.\nonumber\\
& +\kappa_Z m^2_Z \frac{h}{v} Z_{\mu}Z^{\mu}+2\kappa_W m^2_W \frac{h}{v} W_{\mu}W^{\mu} \nonumber\\
 &+ \kappa_\gamma A_{\gamma} \frac{\alpha}{\pi}  \frac{h}{v}F^{\mu\nu}F_{\mu\nu} 
\, , 
\end{align}
where $v=246\,$GeV is the Higgs VEV. The underlying custodial symmetry implies $\kappa_W=\kappa_Z=\kappa_V$, while $\bar\kappa_{qq'}=\bar\kappa_{q'q}^*$ and $\kappa_{V,\gamma,q}$ are real because of the assumed CP conservation. Note that $\bar \kappa_q$ and $\bar \kappa_{qq'}$ are normalized to the SM $b$-quark Yukawa coupling for later convenience.  The $\bar \kappa_{qq'}$ couplings are flavor-violating, while the other couplings are flavor-conserving.
The SM loop function for the $h \gamma\gamma$ coupling is given at one-loop order by 
$A_\gamma \approx -0.81$~\cite{oai:arXiv.org:hep-ph/9505225}. The SM limit corresponds to  $\kappa_\gamma=\kappa_V=1$, and $\bar \kappa_{s}=m_s/m_b\simeq 0.020$, $\bar \kappa_{d}=m_d/m_b\simeq 1.0\cdot 10^{-3}$, $\bar \kappa_{u}=m_u/m_b\simeq 4.7 \cdot 10^{-4}$.  The quark masses are evaluated at  $\mu=m_h$ using NNLO running in the $\overline{\rm MS}$ scheme with low energy inputs from \cite{Beringer:1900zz}.  All of the $\bar \kappa_{qq'}$ vanish in the SM.  Any deviations from these relations would signal the presence of new physics.

\mysection{Constraints from the current data}
In \cite{Delaunay:2013pja} the inclusive production rate at the LHC was used to put an indirect bound on the charm Yukawa coupling.  Here we adapt this analysis to the other Yukawa couplings, $\bar \kappa_i$. The current ATLAS~\cite{ATLASdata}, CMS~\cite{Chatrchyan:2014nva,CMSdata} and Tevatron~\cite{TEVdata} Higgs measurements are included (based on Tables~13 and~14 of Ref.~\cite{Bechtle:2014ewa}), as are the indirect constraints from the LEP electroweak precision measurements \cite{Falkowski:2013dza}. For simplicity, correlations between the different measurements are neglected and asymmetric uncertainties are symmetrized. The quark anti-quark Higgs-fusion cross section is evaluated at next-to-leading order in $\alpha_s$ based on the bottom fusion cross section obtained 
in \cite{Harlander:2003ai} using MSTW parton distribution functions~\cite{Martin:2009iq}.

We begin with the flavor-conserving couplings. A naive $\chi^2$ fit to the data that fixes all Higgs couplings to their SM values, except for one of the up, down, or strange Yukawas at a time, leads to the 95\% confidence level (CL) bounds
\beq
	\left| \bar{\kappa}_u \right| < 0.98  \, , \quad
	\left| \bar{\kappa}_d \right| < 0.93 \,  , \quad
	\left| \bar{\kappa}_s \right| < 0.70 \,  . 
\eeq
If all of the Higgs couplings (including $h\to WW, ZZ, \gamma\gamma, gg, Z\gamma, b\bar{b}$ and $\tau\bar\tau$) are allowed to vary from their SM values, we get the weaker 95\%~CL bounds
\beq
	\left| \bar{\kappa}_u \right| < 1.3  \, , \quad
	\left| \bar{\kappa}_d \right| < 1.4 \,  , \quad
	\left| \bar{\kappa}_s \right| < 1.4 \,  . 	
\eeq

We repeat the analysis for the off-diagonal couplings. The 95\% CL upper bounds obtained when modifying only a single Yukawa coupling at a time (or allowing for modification of the other Higgs couplings as above) are:  
\beq
\left| \bar{\kappa}_{qq'} \right| < 0.6\, (1)\, ,
\eeq
for $q,q'\in u,d,s,c,b $ and $q\neq q'$.  The bounds are 10-20\% stronger for couplings only involving sea quarks, as their slightly smaller direct production cross section does not compensate for the increased decay width.

Inclusive Higgs rate measurements cannot distinguish between the individual $\bar \kappa_{qq'}$. Low energy observables such as neutral meson mixing do place indirect bounds on the individual couplings, with the weakest bound found to be $\left| \bar{\kappa}_{bs} \right|<8\cdot10^{-2}$~\cite{Harnik:2012pb} (see also \cite{Goertz:2014qia}).  However, these bounds are model dependent. For instance, if the Higgs is part of a multiplet that approximately conserves the flavor symmetries, cancellations will occur between the contributions of the Higgs and other members of the multiplet.  The latter could either have reduced production rates or they could mostly decay to light quarks, thus remaining unobserved.

\begin{figure}[!t]
\begin{center}
\includegraphics[width=0.5\textwidth,angle=0]{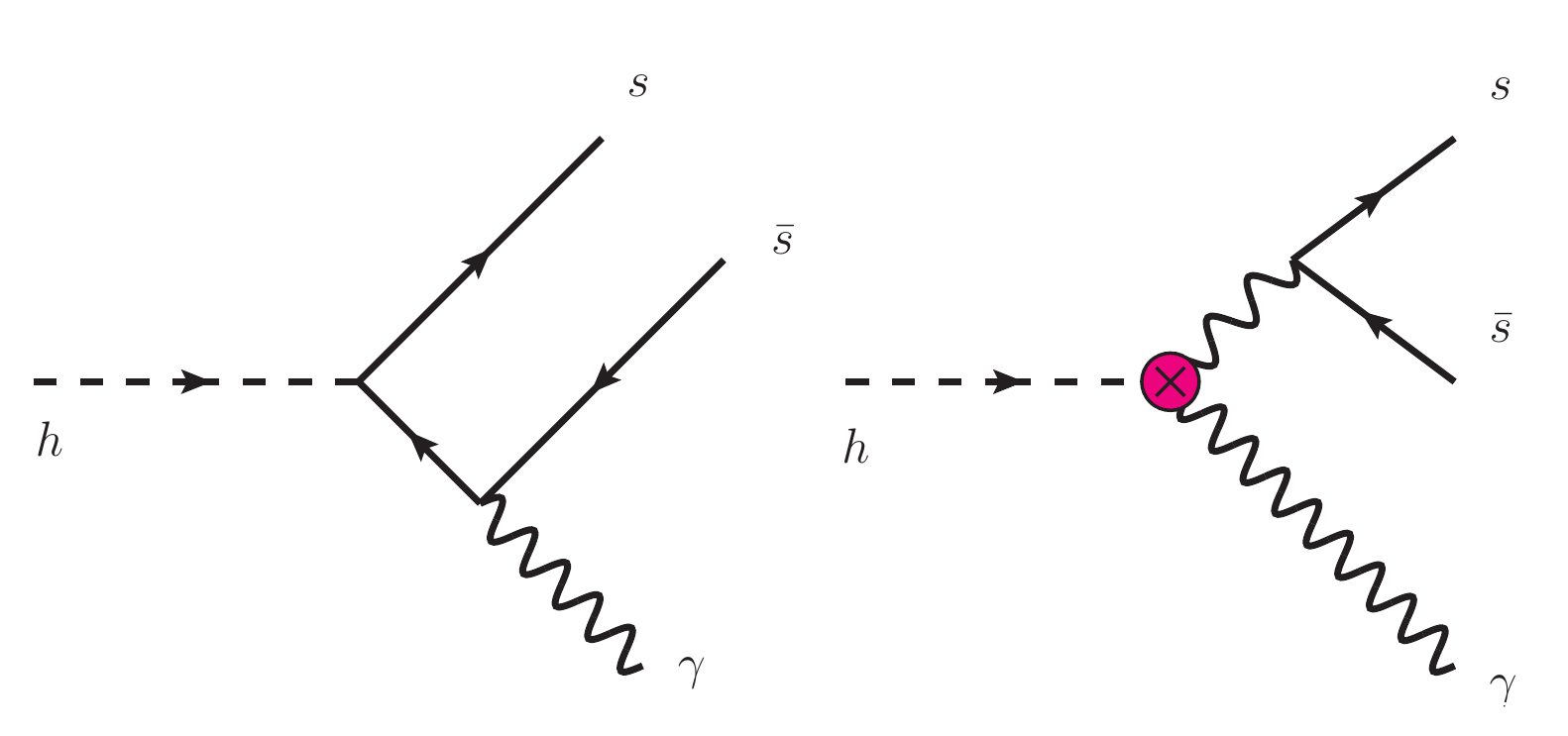}
\end{center}
\caption{Direct-amplitude diagram (left) and indirect-amplitude diagram (right) contributing to $h \to\phi\gamma$.} \label{fig:dd:id}
\end{figure}

\mysection{Flavor-conserving photonic decays}
We begin with $h\to \phi\gamma$.  The decay amplitude receives two dominant contributions which we denote as direct and indirect.  These are shown in Fig. \ref{fig:dd:id}. The indirect contribution proceeds through the $h\gamma\gamma$ coupling, followed by the fragmentation of $\gamma^{*} \to \phi$. The direct amplitude involves a hard $h\to s\bar s \gamma$ vertex, where an intermediate $s$-quark line with an off-shellness $Q^2\sim {\mathcal O}(m_h^2)$ is integrated out.  Its evaluation is a straightforward application of QCD factorization \cite{Beneke:2000ry}. %
The largest sensitivity to the Higgs--strange quark coupling is due to the interference of the two amplitudes which, however, only involves the real part of the coupling, ${\rm Re}(\bar\kappa_s)$. Working in the limit of real $\bar \kappa_s$, the $h\to \phi\gamma$ decay amplitude is 
\beq\label{Mphigamma}
M_{ss}^\phi=\frac{Q_s e}{2}\epsilon^\phi \cdot \epsilon^\gamma \left(\bar {\kappa}_s \frac{m_b}{v} f_\perp^\phi \langle 1/u\bar u \rangle^{\phi}_{\perp}+\frac{4\alpha}{\pi v} \kappa_\gamma A_\gamma \frac{f_\phi m_h^2}{m_\phi}\right),
\eeq
where the first and second terms are the direct and indirect contributions; $f_\perp^\phi$ and $\langle1/u\bar u\rangle^{\phi}_{\perp}$ are the decay constant and inverse moment of the light-cone distribution amplitude (LCDA) defined in Eq.~\eqref{LCDA}, $Q_s e =-e/3 $ is the strange quark electric charge, and $\varepsilon_\gamma$ and $\varepsilon_\phi$ are the $\gamma$ and $\phi$ polarization vectors. We have used the definition $\langle \phi | J_{\rm EM}^{\mu}(0) | 0 \rangle = f_{\phi} m_{\phi} \epsilon_{\phi}^{\mu} $ for the $\phi$ decay constant $f_\phi$, where $J_{\rm EM}^{\mu} = \sum_f Q_f \bar{f} \gamma^{\mu} f$ is the electromagnetic current. Note that for CP violating couplings, $M_{ss}^\phi$ is sensitive to the phase between $A_\gamma$ and $\bar\kappa_\gamma$. 

The LCDA convolution integral is
\beq
{\langle 1/u\bar u\rangle}^{\phi}_{\perp}=\int_0^1du \frac{\phi_{\perp}^\phi(u)}{u(1-u)}.
\eeq
The leading twist chiral-odd LCDA $\phi_{\perp}(u)$ is defined through the following matrix element of the transversely polarized $\phi$ meson on the light-cone \cite{Ball:1998sk,Beneke:2000wa}:
\beq\label{LCDA}
\begin{split}
\langle \phi(p,&\varepsilon_\perp)|\bar s(x) \sigma_{\mu\nu}s(0)|0\rangle =\\
&-i f_\perp^\phi \int_0^1 du e^{iu p\cdot x}(\varepsilon_{\perp\mu}p_\nu -\varepsilon_{\perp \nu}p_\mu)\phi_\perp^\phi(u).
\end{split}
\eeq

The  partial decay width for $h\to \phi \gamma$ decay is 
\beq
\Gamma_{h\to \phi\gamma}=\frac{1}{8\pi} \frac{1}{m_h} |M_{ss}^\phi|^2,
\eeq
where we used the fact that $|\epsilon_\perp^\phi \cdot \epsilon^\gamma|=1$ for the two possible photon polarizations, so that the two corresponding decay amplitudes are equal in size.  The decay widths for $h\to \rho\gamma$ and $h\to \omega \gamma$  are similarly given by
\begin{equation}
\Gamma_{h\to \rho\gamma}=\frac{\left|M_{dd}^\rho- M_{uu}^\rho\right |^2}{16\pi m_h} , \;\;
\Gamma_{h\to \omega\gamma}=\frac{\left|M_{dd}^\omega+ M_{uu}^\omega\right |^2}{16\pi m_h},
\end{equation}
where the amplitudes are obtained from $M_{ss}^\phi$ via the replacements $s\to u,d$ and $\phi \to \rho,\omega$. 
For simplicity we have neglected $\omega-\phi$ mixing.

In our numerical estimates the Gegenbauer polynomial expansions of the $\phi_\perp$ are truncated at second order, yielding
$\langle 1/u\bar u\rangle_\perp^\phi=6.84(42)$, $\langle 1/u\bar u\rangle_\perp^\rho=6.84(36)$, $\langle 1/u\bar u\rangle_\perp^\omega=6.84(72)$,
using the inputs from~\cite{Dimou:2012un} and fixing $\mu=1\,$GeV. 
The decay constants are $f_\phi=0.235(5)\,$GeV, $f_\rho=0.216(6)\,$GeV, $f_\omega=0.187(10)\,$GeV \cite{Dimou:2012un}. We estimate the error on our LO calculation by varying the renormalization scale for $f_\perp^{\phi,\rho,\omega}$ in the range $[0.5,10]\,$GeV.  The variation is combined in quadrature with the  errors quoted in  \cite{Dimou:2012un} to obtain
$f_\perp^\phi=0.191(28)\,{\rm GeV}$, $f_\perp^\rho=0.160(25)\,{\rm GeV}$, $f_\perp^\omega=0.139(27)\,{\rm GeV}$.
Normalizing to the $h\to b\bar b$ branching ratio gives 
\beq
\label{eq:photFC}
\begin{split}
\frac{\BR_{h\to \phi\gamma}}{\BR_{h\to b\bar b}}&=\frac{\kappa_\gamma \big[\big(3.0\pm 0.13)\kappa_\gamma-0.78\bar \kappa_s \big]  \cdot 10^{-6}}{0.57 \bar \kappa_b^2},\\
\frac{\BR_{h\to \rho\gamma}}{\BR_{h\to b\bar b}}&=\frac{\kappa_\gamma \big[(1.9\pm 0.15)\kappa_\gamma-0.24\bar \kappa_u -0.12 \bar \kappa_d\big] \cdot 10^{-5}}{0.57 \bar \kappa_b^2},\\
\frac{\BR_{h\to \omega\gamma}}{\BR_{h\to b\bar b}}&=\frac{\kappa_\gamma \big[(1.6\pm 0.17) \kappa_\gamma-0.59\bar \kappa_u -0.29\bar \kappa_d \big]\cdot 10^{-6}}{0.57 \bar \kappa_b^2},
\end{split}
\eeq
where 
we have neglected the smaller $\bar \kappa_{s,d,u}^2$ terms. The SM $\BR_{h\to b\bar b}=0.57$ is kept explicit in the denominators.  The numerators thus give the $h\to (\phi,\rho,\omega) \gamma$ branching ratios if the Higgs has the SM total decay width.

 The coefficients multiplying $\bar \kappa_{s,d,u}$ have a relative error of ${\mathcal O}(20\%)$.  This means that for $\bar \kappa_i\sim {\mathcal O}(1)$, deviations from the SM predictions for $h\to (\phi,\rho,\omega) \gamma $ can be significantly larger than the present SM errors. The SM errors can be systematically reduced through advances in lattice QCD and 
measurements of the leptonic $\phi, \rho$ and $\omega$ decays.  Moreover, the above predictions are relatively insensitive to potentially more dangerous non-perturbative QCD effects, e.g. power corrections. For instance, the $h\to gg\to g\bar q q\gamma$ transition results in a higher Fock state contribution to $h \to \phi \gamma $ which only enters at the level of a few$\times 10^{-4}$ of the SM $\BR_{h\to \phi\gamma}$. 

\begin{figure}[t]
\centering
\includegraphics[width=3.1in]{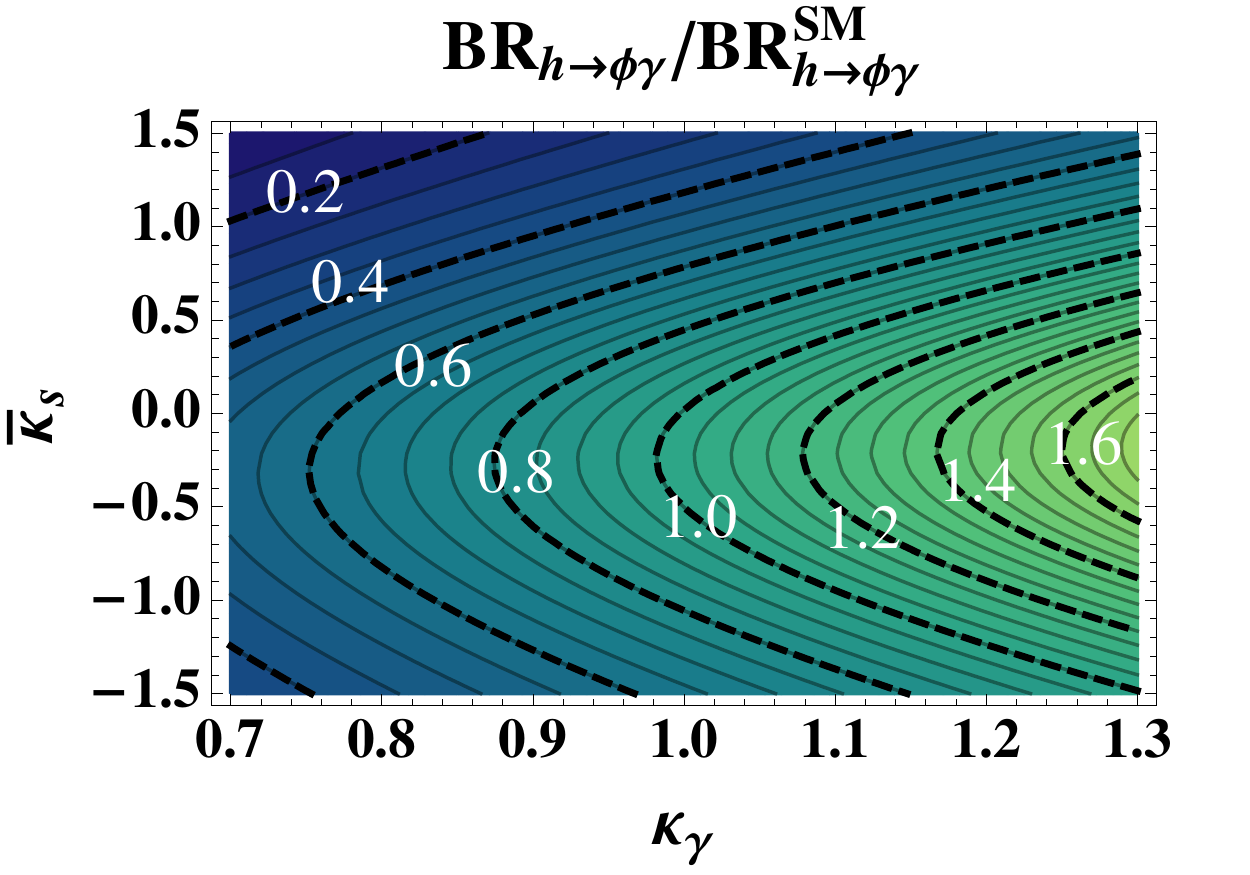}
\caption{The expected deviation in the branching ratio $h \to \phi\gamma$ relative to its SM value as a function of $\kappa_{\gamma}$ and $\bar{\kappa}_s$. } \label{fig:BRdev}
\end{figure}

The expected deviation in the $h\to \phi\gamma$ branching ratio from its SM value is shown in Fig.~\ref{fig:BRdev}, as a function of $\kappa_{\gamma}$ and $\bar \kappa_s$. We note that the direct amplitude by itself contributes at the ${\mathcal O}(10^{-11})$ level in the SM.  Only the interference with the indirect term allows this mode to be potentially observable. 

\mysection{Flavor-violating photonic decays}
The radiative decays $h\to V\gamma$, where $V=B^{*0}_s, \,B^{*0}_d,\, K^{*0}, D^{*0}$ provide interesting possibilities to probe the flavor-violating Higgs couplings $\bar\kappa_{bs,sb}$, $\bar\kappa_{bd,db}$, $\bar\kappa_{ sd,ds}$ and $\bar\kappa_{cu,uc}$.  These flavor-violating decays only receive direct amplitude contributions, since photon splitting preserves flavor. The $h\to K^{*0}\gamma$ rate is readily obtained from the results of the previous section, yielding an $\mathcal{O}(10^{-8} )$ branching ratio for $\bar \kappa_{ds}\sim {\mathcal O}(1)$, out of reach of planned colliders. We thus focus on the decays to heavy mesons. 

The essential difference with respect to the light mesons is that the $B_{(s)}^{*0}$ and $D^{*0}$ LCDA are heavily weighted toward the $b$ and $c$ quarks (we treat the $c$ quark as heavy, $m_c\gg \Lambda_{\rm QCD}$).  Focusing first on the $h\to \bar B_s^{*0}\gamma$ decay, 
the dominant contribution comes from the diagram where the photon is emitted from the $s$-quark intermediate leg. Emission of the photon from an intermediate $b$ quark line is ${\mathcal O}(\Lambda_{\rm QCD}/m_b)$ suppressed and can be neglected. We thus obtain the partial decay width
\beq
\Gamma_{h\to \bar B_s^{*0}\gamma}=\frac{1}{8\pi}\frac{1}{m_h}\left(\frac{f_{B_s} m_{B_s}}{2}\frac{m_b}{v}   \frac{Q_s e_0}{\lambda_B(\mu)}\right)^2 \frac{|\bar \kappa_{bs}|^2+|\bar \kappa_{sb}|^2}{2},
\eeq
where HQET sum rule estimates of the inverse moment of the $B$ meson LCDA yield $\lambda_B(\mu)=(460\pm110)\,$MeV for $\mu=1\,$GeV \cite{Braun:2003wx} (see also~\cite{Khodjamirian:2013iaa}). Note that $\lambda_B$
can be determined from $B\to \ell\nu \gamma$.  Present limits, including NLO radiative corrections, yield a result compatible with the above estimate \cite{Beneke:2011nf,Braun:2012kp}.  We have assumed flavor SU(3) symmetry so that $\lambda_B$ is the same for $B^0_s$ and $B^0_d$. Numerically one has, 
\begin{align}
\frac{\BR_{h\to \bar B_s^{*0}\gamma}}{\BR_{h\to b\bar b}}&=\frac{\BR_{ \bar B_s^{*0}\gamma}^{(1)}}{0.57 \bar \kappa_{b}^2}\frac{ |\bar\kappa_{bs}|^2+|\bar \kappa_{sb}|^2}{2},
\end{align}
where $\BR_{\bar B_s^{*0}\gamma}^{(1)}=(2.1\pm 1.0) \cdot 10^{-7}$. The $h\to \bar B^{*0}\gamma$ and $h\to D^{*0}\gamma$ branching ratios are obtained by replacing $\bar \kappa_{bs,sb}$ with $\bar \kappa_{bd,db}$ and  $\bar \kappa_{cu,uc}$, respectively, with $\BR_{\bar B^{*0}\gamma}^{(1)}=(1.4\pm 0.7) \cdot 10^{-7}$ and $\BR_{D^{*0}\gamma}^{(1)}=(8.6\pm 8.3) \cdot 10^{-8}$. We have taken $\lambda_D=\lambda_B$, but have inflated the errors on $\lambda_D$ by a factor of 2. For the decay constants we have used the FLAG averages $f_{B_s}=228(5)\,$MeV, $f_{B}=191(4)\,$MeV, and $f_{D}=209(3)\,$MeV~\cite{Aoki:2013ldr}.  

The $B^{*0}$ and $B_s^{*0}$ decay rates are large enough to be potentially observable at future high-luminosity hadron colliders, see below.  These decay modes are negligible in the SM, where $\bar \kappa_{ij}=0$.  Thus, their observation would provide definitive evidence for new physics in the Higgs Yukawa sector.

\mysection{Exclusive decays with $W$ and $Z$}
The charged $h\to M^-W^+$ decays differ qualitatively from the radiative decays. Because the $W$ attaches itself to a charged current one can probe flavor violating couplings of the higgs involving top quarks. The complication is that the $W$ can have both transverse and longitudinal polarizations, yielding lengthier analytical expressions that will be presented elsewhere~\cite{future}. Numerically, we find for the  most promising mode
\beq
\begin{split}
\frac{\BR_{h\to B^{*-} W^+}}{\BR_{h\to b\bar b}}&\simeq\frac{1.2 \cdot 10^{-10}\big[ \kappa_{V}^2+22 \bar \kappa_{tu}^2+26 \bar \kappa_{ut}^2+\cdots\big]}
{0.57 \bar \kappa_{b}^2},
\end{split}
\eeq
where only the potentially largest contributions are shown.  The bounds on $\bar\kappa_{tu,ut}$ and $\bar \kappa_{tc,ct}$ from $t\to hu, hc$ decays are \cite{Aad:2014dya,CMStop} $(|\bar\kappa_{tc}|^2+|\bar\kappa_{ct}|^2+|\bar \kappa_{tu}|^2+|\bar\kappa_{ut}^2|)^{1/2} <7.1 \, ,$ which implies that $\BR_{h\to B^{*-} W^+} \le1.6 \cdot 10^{-7}$ is allowed. 

Decays of the form $h\to MZ$ are closer to the radiative decays discussed above, since they only involve neutral currents. An important difference is that the interference terms between the direct and indirect amplitudes are smaller.  Thus, these decays are less useful for measuring the Higgs couplings to light quarks. 

\mysection{Future experimental perspectives}
We begin our discussion of the experimental prospects for these decays by estimating the number of events expected at future collider facilities.  We focus on the $h \to \phi\gamma$ mode and use {\tt Pythia~8.1}~\cite{Sjostrand:2007gs} to estimate its main features in proton-proton collisions  at the LHC with the 
center of-mass energy of\,14 TeV. The main $\phi$ decay modes ($K_{L,S}$, $K^{\pm}$, $\pi^{\pm}$ and $\pi^0$) were explored.
 In  $70$ to $75\,$\% cases the kaons/pions and the prompt photon have $|\eta|<2.4$  and are thus within the minimal fiducial volume of the 
ATLAS and CMS experiments.  We therefore adopt the geometrical acceptance factor of $A_{\rm g}=0.75$ below, but do not include other efficiency or trigger factors. 

We focus on the following three facilities which were considered in the Snowmass Higgs working group~\cite{Dawson:2013bba}: a HL-LHC, a high-energy LHC (HE-LHC), and a VLHC.  The Higgs production cross sections at these machines are obtained from the LHC Higgs cross section working group~\cite{lhcxs}.  We have assumed two detectors for the HL-LHC and a single detector for the other colliders.  

We estimate the reach in $\bar \kappa_s$ that can be obtained, given the current theoretical uncertainties and the expected statistical errors. For simplicity, we assume $\kappa_\gamma =1$ as in the SM. The significance of a deviation in the measured value of $\BR_{h\to\phi\gamma}$ with respect to its SM prediction can be quantified by
${\cal S} = | \BR_{h\to\phi\gamma} - \BR_{h\to\phi\gamma}^{\rm SM} | / (\delta\BR_{h\to\phi\gamma})$,
where
$ \left( \delta\BR_{h\to\phi\gamma} \right)^2
 =   \BR_{h\to\phi\gamma} /(\sigma_h {\cal L}  A_{\rm g} )  +  ( \delta\BR^{\rm th}_{h\to\phi\gamma} )^2\, $
 is the estimated uncertainty. The first term is the statistical uncertainty ($\sigma_h$ is the total Higgs production cross section and ${\cal L}$ is the integrated luminosity), while the second term is the theoretical one, $\delta\BR^{\rm th}_{h\to\phi\gamma}\approx1.3\cdot 10^{-7}$ for $\kappa_\gamma=1$, see Eq.~(\ref{eq:photFC}).  Our criterion for a large-enough deviation from the SM prediction is ${\cal S}\ge3$. Our results are summarized in Table \ref{tab:rates}.
\begin{table*}[t]
\begin{tabular}[t]{ccccc}
\hline\hline
~~$\sqrt{s}\,$[TeV]~~ & ~~$\int {\cal L}\, dt\,$[fb$^{-1}$]~~ & ~~\# of events (SM) ~~ & ~~~$\bar \kappa_s>(<)$~~~ & ~~~$\bar \kappa_s^{\rm stat.}>(<) $ ~~~\\
\hline
$14$ & $3000$ & $770$ & $0.39\,(-0.97)$ & $0.27\,(-0.81)$  \\
$33$ &  $3000$ & $1380$  & $0.36\,(-0.94)$ & $0.22\,(-0.75)$ \\
$100$  & $3000$ & $5920$ &  $0.34\,(-0.90)$ &  $0.13\,(-0.63)$ \\
\hline\hline
\end{tabular}
\caption{Three future hadron colliders with expected center of mass energies, integrated luminosities,  number of $h\to\phi\gamma$ events,  the minimal (maximal) values of $\bar \kappa_s$ that can be probed with present (4th column) and negligible (last column) theory error, see text.
}
\label{tab:rates}
\end{table*}

We note that only a few events are expected in electron-positron colliders (ILC, ILC with luminosity upgrade, CLIC), probably too small to allow for observation. Although $\approx 30$ events are expected at the TLEP collider, this is still too small to probe a significant deviation from the SM prediction.  Thus, the possibility of observing this mode appears to be unique to future hadron machines.

The $h\to\phi\gamma$ mode offers several promising experimental handles. The decay products, kaons or pions, fly in a narrow cone, $\Delta R<0.1$, with tens of GeV of energy.  They reach the detector before they decay (except the $K_S$ and $\pi^0$, which have much shorter lifetimes). The most apparent features for identification of the charged decay modes are the near collinearity of the photon and the $\phi$-jet in the transverse plane, the jet sub-structure information (two close high-$p_T$ tracks in a narrow cone) and the di-track invariant mass distribution assuming kaons/pions. A detailed experimental simulation will be required to determine if this signature is feasible.  

The $h \to\rho\gamma$ and $h\to\omega\gamma$ modes have rates comparable to or larger than the $\phi$ channel, see Eq.~(\ref{eq:photFC}).  The $\rho$ decays almost exclusively to $\pi^+ \pi^-$.  This is a relatively clean mode, similar to the $\phi \to K^+ K^-$ decay, and features two tracks with high transverse momenta and a proper invariant mass.  The $\omega$ decays to $\pi^+\pi^-\pi^0$.  This will be harder to trigger on than the $\rho$ or $\phi$ modes, as the transverse momenta of the charged pions are lower and the hard-to-identify neutral pion smears the observable quantities.  A detailed experimental study is required to assess the feasibility of this channel.  The  $h\to \bar B^{*0}\gamma$ mode will be more difficult, as the $B^{*0}$ decays to $B^{0} \gamma$, leading to a $b$-jet + $\gamma$ final state.  More study of this mode is needed.
  
\mysection{Conclusions}
In this Letter we have shown that rare Higgs decays to vector mesons can explore the 
structure of the Higgs Yukawa couplings to the first and second generation quarks.  Directly accessing the 
couplings of the Higgs to the lightest quarks was previously thought to be impossible.  
Rare decays of the form $h \to MV$
offer sensitivity to both flavor-conserving and flavor-violating couplings of the Higgs.  They are theoretically calculable, 
experimentally promising, and should become a priority at the LHC Run II and at future hadron colliders.  We look 
forward to further investigation of the ideas we have discussed here.
 
\mysection{Acknowledgements}
 We thank the CERN theory group, where this work was initiated, for its hospitality.  F.~P. thanks K.~Mishra for helpful discussions.  The work of A.~K. is supported by the DOE grant DE-SC0011784.  G.~P. is supported by the Minerva foundation, the IRG and by the Gruber award.  F.~P. is supported by the DOE grants DE-FG02-91ER40684 and DE-AC02-06CH11357.  S.~S.  is supported by the DOE grant DE-FG02-91ER40684.  J.~Z. is supported in part by the U.S. National Science Foundation under CAREER Grant PHY-1151392.

\end{document}